# COPS AND INVISIBLE ROBBERS:
# THE COST OF DRUNKENNESS

ATHANASIOS KEHAGIAS, DIETER MITSCHE, AND PAWEŁ PRAŁAT

ABSTRACT. We examine a version of the *Cops and Robber* (CR) game in which the robber is *invisible*, i.e., the cops do not know his location until they capture him. Apparently this game (CiR) has received little attention in the CR literature. We examine two variants: in the first the robber is *adversarial* (he actively tries to avoid capture); in the second he is *drunk* (he performs a random walk). Our goal in this paper is to study the *invisible Cost of Drunkenness* (iCOD), which is defined as the ratio $ct_i(G)/dct_i(G)$, with $ct_i(G)$ and $dct_i(G)$ being the *expected capture times* in the adversarial and drunk CiR variants, respectively. We show that these capture times are well defined, using game theory for the adversarial case and partially observable Markov decision processes (POMDP) for the drunk case. We give exact asymptotic values of iCOD for several special graph families such as $d$-regular trees, give some bounds for grids, and provide general upper and lower bounds for general classes of graphs. We also give an infinite family of graphs showing that iCOD can be arbitrarily close to any value in $[2, \infty)$. Finally, we briefly examine one more CiR variant, in which the robber is invisible and "*infinitely fast*"; we argue that this variant is significantly different from the *Graph Search* game, despite several similarities between the two games.

## 1. INTRODUCTION

*Cops and Robber* (CR) is a game introduced by Nowakowski and Winkler [38] and (independently) by Quilliot [44]. CR is played on a fixed, undirected, simple and finite graph $G$ where $K$ cops pursue a single *robber*. Both the cops and the robber are located in the vertices of $G$ and everybody has full information about everybody else's current location. In every turn of the game, first the cops and then the robber can move to their new locations along the edges of the $G$. The cops win if they *capture* the robber, i.e., if at least one cop is in the same vertex as the robber; the robber's goal is to avoid capture. This is the "classical", extensively studied version of the CR game. Many other versions have been proposed and studied; for a review of the literature see [2, 6, 19] and the book [5].

We call the robber of the classical CR game *adversarial*: he wants to avoid capture for as long as possible and plays optimally towards this end. In a recent paper [26] we have studied a CR variant where the robber is *drunk*, i.e., he performs a random walk on $G$ (he does not attempt to avoid capture; essentially he is oblivious of the cops). This is really a one-player game. In [26] we have compared the adversarial and drunk CR variants and have especially studied the *Cost of Drunkenness* (COD), i.e., the ratio of "adversarial" capture time and "drunk" expected capture times.

In the current paper we are concerned with the much less studied version of the CR game, in which the robber is *invisible* to the cops (unless they are located in the same vertex). All the other rules remain the same as in the classical CR and we examine both the adversarial and the

---

drunk variants. We will call this game cops-and-*invisible*-robber (*CiR*). Our main goal is (as in [26]) to study the "invisible COD" (iCOD).

CiR and, more generally, invisible robber versions of CR have so far received little attention in the graph theory literature. To the best of our knowledge, the first paper that deals with the invisible robber is [52]. More recent work in which the cops have either zero, incomplete or intermittent information about the robber's location include [12, 13, 25, 51]. In addition, some works [11, 14, 15, 16] examine the case in which the cops' incomplete observations are augmented by the use of traps, alarms, radars and so on. The variant of an intermittently observed robber is related to another variant, where the robber moves with greater speed than the cops; this is studied in [7]. All these works share a common approach, according to which cops use a *predetermined* move sequence by which capture is guaranteed. Because the cops will only see the robber when they capture him, this sequence does not depend on actual robber moves, hence it can be computed before the CiR game starts. Furthermore, the robber can also compute the same sequence, so he is *omniscient* in the sense that he knows all cop moves before the game starts. This approach is also used in *graph search* (GS) games in which the robber is also endowed with infinite speed. However, we will argue in Section 7 that GS is a different game from CiR (the seminal paper for graph search is [42]; good recent reviews are [2, 6, 19]).

In this paper we are interested in a different approach, which improves the cops' fortunes. Since the cops will never see the robber, they must predetermine their *strategy*; but this can be *randomized*, so the actual cop *moves* do not need to be predetermined at the beginning of the game. Hence, in CiR, at every time step the robber will know *previous* cop moves but not the future ones. As we will show in later sections, the use of randomized strategies can, in some cases, reduce the number of cops necessary to capture the robber, provided that our goal is to obtain a finite expected capture time and so cops win the game after a finite number of steps with probability one.

Randomized strategies have not received much attention in the graph theoretic CR literature. From the few papers which explore this approach, we mention [1, 23, 24] (note that in [1] *both* cops and robber are invisible to each other until they occupy the same vertex). Brief mentions of randomized strategies also appear in [11, 51].

On the other hand, there is a large *robotics* literature which deals with the (more general) *pursuit/evasion* problem; while roboticists do not often use the term "cops and robber", they have studied pursuit/evasion on graphs using formulations quite similar to CiR, especially for the case of the drunk robber; see for example [8, 9, 21, 22, 28, 29, 39, 53, 55] and the review [10]. As expected, this research is more application-oriented. Also, the *operations research* community has studied what is essentially the search for an invisible drunk robber in a graph; we only cite here three representative works here [29, 49, 50].

Most of the works cited in the previous paragraph handle the invisible drunk robber with tools from the theory of *partially observable Markov decision processes* (POMDP; see [31, 35, 48]) and this is the approach we will use in the current paper. For the invisible adversarial robber, we believe the "natural" treatment is through game theoretic methods; in fact what is required is the generalization of POMDP's to *stochastic* or *Markovian games*. The subject was introduced in [46]; some of the subsequent results can be found in [3, 17, 18, 27, 33, 34, 36, 37, 40, 41, 45, 54]; a paper we have found especially useful is [20].

The rest of the paper is organized as follows. In Section 2 we introduce preliminary notation and results. In Section 3 we give an extended example which illustrates the basic aspects of the CiR game and, perhaps more importantly, establishes that the iCOD can become arbitrarily large. In Section 4 we prove that the value of the CiR game always exists (for both the adversarial and drunk variant). In Section 5 we provide general upper and lower bounds, which are then used and tightened in Section 6 to obtain (almost) exact values for special graph classes, among them $d$-regular trees and grids; we also introduce the family of *broom* graphs and use it to show that the iCOD can be arbitrarily close to any value in $[2, \infty)$. In Section 7 we briefly study CiR (and compare it to GS) when the robber has "infinite" speed. In Section 8 we summarize and discuss our conclusions.

## 2. Preliminaries

Let $G = (V, E)$ be a fixed undirected, simple, connected and finite graph. We begin by listing definitions, notation, assumptions and a useful theorem.

(1) $\mathbb{N} = \{1, 2, 3, \ldots\}$ and $\mathbb{N}_0 = \{0, 1, 2, \ldots\}$.
(2) We will always use $n$ to denote the number of vertices of $G$ (i.e., $|V(G)| = n$).
(3) We assume that the cops-and-robber game (both the CR and CiR versions) are played by two players, denoted by C (the cop player) and R (the robber player).
(4) C controls $K$ cops ($K \in \mathbb{N}$, we will denote $\{1, 2, \ldots, K\}$ by $[K]$). $X_t^k \in V$ denotes the position of the $k$-th cop at time $t$ ($k \in [K]$, $t \in \mathbb{N}_0$); $X_t = (X_t^1, X_t^2, \ldots, X_t^K) \in V^K$ denotes the vector of all cop positions at time $t$.
(5) R controls a single robber, whose position at time $t \in \mathbb{N}_0$ is denoted by $Y_t \in V$.
(6) Actually R is active only in the adversarial variants. In the drunk variants the robber performs random walk on $G$ (we could say he is controlled by *Chance*) according to the following rules

$$\forall u \in V, \Pr(Y_0 = u) = \frac{1}{|V|} \tag{1}$$

$$\forall u \in V, \Pr(Y_{t+1} = u \mid Y_t = v) = \begin{cases} \frac{1}{|N(v)|} & \text{if } u \in N(v) \\ 0 & \text{otherwise.} \end{cases} \tag{2}$$

($N(v)$ is the *neighbourhood* of vertex $v$; $N(v)$ does *not* include $v$).
(7) The moving sequence in the adversarial variants is as follows. At $t = 0$, first C chooses initial positions $X_0 \in V^K$, then R chooses $Y_0 \in V$. For $t \in \mathbb{N}$ first C chooses $X_t \in V^K$ and then R chooses $Y_t \in V$. Movements are possible only along edges of $G$; that is, $\{X_t^k, X_{t+1}^k\} \in E$ and $\{Y_t, Y_{t+1}\} \in E$, for all $k \in [K]$ and $t \in \mathbb{N}_0$.
(8) The moving sequence in the drunk variants is the same, except that $Y_0, Y_1, Y_2, \ldots$ are chosen by Chance, according to equations (1)-(2).
(9) The *capture time* is denoted by $T$ and defined as follows

$$T = \min\{t : \exists k \in [K] \text{ such that } X_t^k = Y_t\};$$

that is, $T$ is the first time a cop is located at the same vertex as the robber (note that this can happen during either a cop move or a robber move). If capture never takes place, then $T = \infty$. As will be seen in the sequel, capture time will in general be a random variable, dependent on the *strategies* used by the cop and the robber (and also on $K$).

In what follows, unless stated otherwise, it will be assumed that (a) C's goal is to capture the robber as fast as possible and (b) he plays optimally with respect to this goal; we will summarize these assumptions by saying that C is *adversarial*. R can be in one of two modes: *adversarial* (he plays optimally to avoid capture for as long as possible) or *drunk* (he simply performs a random walk on $G$). The cops' locations are always known to the adversarial R (specifically, at time $t$ he knows $X_0, X_1, \ldots, X_t$). The robber can be *visible* (his location is known to the cops) or *invisible* (his location is unknown).

When the robber is *visible* and *adversarial*, $T$ is deterministic [26]. The *cop number* of $G$ is denoted by $c(G)$ and defined to be the minimum $K$ for which $T < \infty$ (there is always such a $K$, less than or equal to $|V|$). We define $ct(G)$ to be the "optimal capture time given that $K = c(G)$ and C and R play optimally" (this definition requires some care, see [26]). When the robber is *visible* and *drunk*, he performs a random walk on $G$ as indicated by (1)-(2). In [26] we show that the following quantity is well defined

$$dct(G) = E(T \mid \text{ when } K = c(G), \text{ C plays optimally, R is drunk}).$$

Hence the *cost of (visible) drunkenness* is also well defined by

$$F(G) = \frac{ct(G)}{dct(G)},$$

and we obviously have $F(G) \geq 1$ (capturing the adversarial robber is at least as hard as capturing the drunk one, since the former can always choose to behave as if he were drunk).

Let us now turn to the *invisible* robber. In [26] we have proved the following.

**Theorem 2.1.** *Suppose that $c(G)$ cops perform a random walk on a connected graph $G$, starting from any initial position and the robber is adversarial. Then $E(T) < \infty$.*

It follows that $c(G)$ *adversarial* cops suffice to capture the invisible adversarial robber. On the other hand, capturing the invisible robber is at least as hard as capturing the visible one, hence $c(G)$ is also the minimum required number of cops. In short, *the cop number of a graph is the same for the visible and invisible CR version* (however, the expected capture time $T$ will generally be bigger in the invisible variant, comparing to the capture time in the visible one). In the rest of the paper we will always assume that the number of cops is $K = c(G)$.

For the invisible variant of the game we will define

$$ct_i(G) = E(T \mid K = c(G), \text{ C and R play optimally}), \tag{3}$$

$$dct_i(G) = E(T \mid K = c(G), \text{ C plays optimally, R is drunk}), \tag{4}$$

$$F_i(G) = \frac{ct_i(G)}{dct_i(G)} \geq 1. \tag{5}$$

$F_i(G)$ is the "*invisible cost of drunkenness*" (iCOD). Note that we will always assume that $n \geq 2$ and so $ct_i(G) \geq dct_i(G) > 0$. The existence of the quantities defined in (3)-(5) and the meaning of "optimal play" require careful study, which will be deferred to Section 4; we will first present an extended example in Section 3.

## 3. The Invisible Cost of Drunkenness Can be Arbitrarily Large

To clarify some of the key CiR concepts, we now present an extended example which involves the $N$-star graphs $S_N$, illustrated in Figure 1 and defined as follows (note that, for all $N$, we have $n = |V| = N + 1$). For $N \in \mathbb{N}$, $S_N$ has vertex set $V = \{0, 1, \ldots, N\}$ and edge set $E = \{(0, 1), (0, 2), \ldots, (0, N)\}$. Obviously a single cop can catch the *visible* robber in $S_N$. So we will study CiR with a single cop, as well.

A robber *strategy* generates a robber's move for every time $t$, based on the information available to R at time $t$; this information is the cop moves $X_0, X_1, \ldots, X_t$ and the robber moves $Y_0, Y_1, \ldots, Y_{t-1}$. As will become clear, R may gain an advantage by *randomizing* his strategies. Hence a robber strategy $\sigma_R$ is an infinite sequence of probability distributions conditioned under previous moves/positions of both cops and robber:

$$\sigma_R = \{\Pr\left(Y_t \mid X_0 = x_0, \ldots, X_t = x_t, Y_0 = y_0, \ldots, Y_{t-1} = y_{t-1}\right)\}_{t=0}^{\infty}. \tag{6}$$

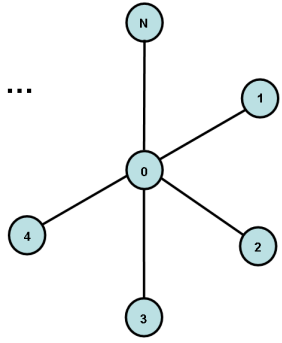

FIGURE 1. The $N$-star $S_N$.

The strategies must also be *feasible*, i.e., only moves along edges can have positive probabilities. Similarly a cop strategy $\sigma_C$ is an infinite sequence of probability distributions conditioned under previous moves/positions of *only* the cops (and an initial probability, for $t = 0$):

$$\sigma_C = \{\Pr\left(X_t \mid X_0 = x_0, \ldots, X_{t-1} = x_{t-1}\right)\}_{t=1}^{\infty} \cup \{\Pr\left(X_0\right)\}. \tag{7}$$

The conditionals must satisfy the feasibility requirement. Note that conditioning is on cop moves only, since the robber is invisible to the cops. However, let us stress that the sequence of previous moves of the cops affects their strategy for the future, since the fact that the robber was not seen at these vertices before contains important information. In the adversarial variant, the expected capture time depends on both $\sigma_R$ and $\sigma_C$, hence we will write $E\left(T \mid \sigma_R, \sigma_C\right)$. In the drunk variant we will write $E\left(T \mid \sigma_C\right)$ instead.

Now let us consider optimal cop and robber strategies for $S_N$. We will first deal with the adversarial variant. Let us fix some $N \geq 2$ (so $S_N$ is a tree with $N$ leaves) and let C use the following strategy: he chooses a permutation $u_1 u_2 \ldots u_N$ of the set $[N]$ with uniform probability $\left(\Pr\left(u_1 u_2 \ldots u_N\right) = \frac{1}{N!}\right)$ and the cop moves as follows

$$X_0 = u_1, X_1 = 0, X_2 = u_2, X_3 = 0, X_4 = u_3, \ldots, X_{2N-2} = u_N.$$

In other words, the cop starts at a random leaf and visits every other leaf in a random order (without repetitions). Therefore there exists a cop strategy $\widehat{\sigma}_C$ of the form (7) which produces this sequence.

Now, R sees $X_0 = u_1$ before placing the robber and he knows that the cop has no incentive to stay in place, so he infers that $X_1 = 0$. Hence R knows that $Y_0$ must belong to $\{u_2, u_3, \ldots, u_N\}$ but he has no incentive to prefer one of these (since he is unaware of the order by which they will be visited by the cop), and thus he will place the robber equiprobably in one of $\{u_2, u_3, \ldots, u_N\}$. After the initial placement, R will never move the robber because the only possible move is into 0 and this would result in a capture (either the robber runs into the cop during an odd turn, or vice versa during an even turn). Hence the robber will use the strategy $\widehat{\sigma}_R$ which sets $Y_0 = v \in \{u_2, u_3, \ldots, u_N\}$ with probability $\frac{1}{N-1}$, and $Y_t = Y_0$ for $t = \mathbb{N}$. For every initial cop placement we can compute

$$E\left(T \mid \widehat{\sigma}_R, \widehat{\sigma}_C, X_0 = u_1\right) = \frac{1}{N-1} \cdot 2 + \frac{1}{N-1} \cdot 4 + \ldots + \frac{1}{N-1} \cdot (2N-2) = \frac{2}{N-1} \cdot \frac{(N-1) \cdot N}{2} = N$$

and, since there are $N$ equiprobable choices for $u_1$, we also have

$$E\left(T \mid \widehat{\sigma}_R, \widehat{\sigma}_C\right) = N.$$

As we have already argued, $\widehat{\sigma}_R$ gives to R the best possible result (longest expected capture time) given C uses $\widehat{\sigma}_C$. In other words,

$$\max_{\sigma_R} E\left(T \mid \sigma_R, \widehat{\sigma}_C\right) = E\left(T \mid \widehat{\sigma}_R, \widehat{\sigma}_C\right) = N$$

from which follows

$$\min_{\sigma_C} \max_{\sigma_R} E\left(T \mid \sigma_R, \sigma_C\right) \leq E\left(T \mid \widehat{\sigma}_R, \widehat{\sigma}_C\right) = N. \tag{8}$$

By the same argument, if the cop decided to start at the root and visit the leaves in a randomly chosen order, the expected capture time of the best robber (choosing uniformly at random any leaf) would also be $N$. On the other hand, suppose R uses $\widehat{\sigma}_R$ irrespective of C's strategy. Clearly C must use a strategy which visits every vertex and, since he has no reason to prefer a particular order of visitation, he will do just as well by using $\widehat{\sigma}_C$ (he gains no advantage by visiting the same vertex twice). Hence we have

$$N = E\left(T \mid \widehat{\sigma}_R, \widehat{\sigma}_C\right) = \min_{\sigma_C} E\left(T \mid \widehat{\sigma}_R, \sigma_C\right) \Rightarrow$$

$$N = E\left(T \mid \widehat{\sigma}_R, \widehat{\sigma}_C\right) \leq \max_{\sigma_R} \min_{\sigma_C} E\left(T \mid \sigma_R, \sigma_C\right). \tag{9}$$

Finally, we know that

$$\max_{\sigma_R} \min_{\sigma_C} E\left(T \mid \sigma_R, \sigma_C\right) \leq \min_{\sigma_C} \max_{\sigma_R} E\left(T \mid \sigma_R, \sigma_C\right). \tag{10}$$

From (8)-(10) we see that

$$\max_{\sigma_R} \min_{\sigma_C} E\left(T \mid \sigma_R, \sigma_C\right) = E\left(T | \widehat{\sigma}_R, \widehat{\sigma}_C\right) = \min_{\sigma_C} \max_{\sigma_R} E\left(T \mid \sigma_R, \sigma_C\right) = N.$$

In other words, $\widehat{\sigma}_R$ and $\widehat{\sigma}_C$ are optimal strategies for R and C, respectively, and the optimal expected capture time in the adversarial variant is

$$ct_i\left(S_N\right) = N = n - 1, \tag{11}$$

for all $N \geq 2$; it is easy to see that (11) also holds for $N = 1$.

Let us now turn to the drunk variant. In this case only C has to choose a strategy and the optimal $\widetilde{\sigma}_C$ is obvious: he places the cop at $u = 0$ and just waits there. For any $N \geq 1$, the robber will start at 0 with probability $\frac{1}{N+1}$ or at some $u \in \{1, 2, \ldots, N\}$ with probability $\frac{N}{N+1}$. In the former case $T = 0$; in the latter case the robber will move into 0 at $t = 1$, resulting at $T = 1$. Hence

$$dct_i\left(S_N\right) = E\left(T \mid \widetilde{\sigma}_C\right) = \frac{N}{N+1} = \frac{n-1}{n}. \tag{12}$$

From (11) and (12) we get $F_i\left(S_N\right) = N + 1 = n$. It follows that the cost of drunkenness $F_i(S_N)$ can attain any integer in $\mathbb{N}$. Our results are summarized in the following.

**Theorem 3.1.** *For every $N \geq 1$ we have*

$$ct_i(S_N) = N = n - 1, \qquad dct_i(S_N) = \frac{N}{N+1} = \frac{n-1}{n}, \qquad F_i(S_N) = N + 1 = n.$$

## 4. The Cost of Drunkenness is Well Defined

In the previous section we have proved that $F_i(G)$ can take arbitrarily large values. Our proof involved only a particular family of graphs, the $N$-stars. In this section we will show that iCOD is well defined for *every* graph $G$. The real issue, of course, is whether $ct_i(G)$ and $dct_i(G)$ are well defined. Settling this question requires the use of game theoretic concepts for $ct_i(G)$ and POMDP concepts for $dct_i(G)$. For simplicity, we will consider the case of a single cop ($K = 1$); the generalization to $K > 1$ is straightforward. Hence we pick a graph $G$ with $c(G) = 1$ and keep it fixed for the rest of the section (but Lemma 4.1 and Theorems 4.2 and 4.4 remain true for any $G$).

### 4.1. Adversarial Robber.
Here we show that $ct_i(G)$ is well-defined. Our notation and analysis follows very closely [20]. The adversarial variant of CiR is a two player game, which we will call $\Gamma$, played on $G$. Note that $\Gamma$ can last an infinite number of turns (the robber is never captured). We will also make use of auxiliary, *truncated games*: $\Gamma_m$ is the same game as $\Gamma$ but is played for a maximum of $m$ turns.

C moves the cop according to a strategy $\sigma_C$. For a rigorous definition of strategy we need the following.

(1) $A_C = V$ is the set of possible C actions (possible placement of the cop on $G$).
(2) $H_C^{(m)} \subseteq V^m$ is the set of *feasible $m$ long sequences* of cop configurations.
(3) $H_C = \bigcup_{m=0}^{\infty} H_C^{(m)}$ is the set of all *finite-length* feasible cop histories.
(4) $\mathbf{P}(A_C)$ is the set of all probability functions on C actions.

Hence a cop strategy is a function $\sigma_C : H_C \to \mathbf{P}(A_C)$, i.e., a function which maps to every finite-length history $x_0, x_1, \ldots, x_{t-1}$ a probability (conditional on $x_0, x_1, \ldots, x_{t-1}$ when $t \geq 0$) on the next C move $X_t$. We are interested in *feasible* strategies, i.e., those which assign positive probabilities only to feasible next-step cop configurations; we will denote the set of all feasible C strategies by $\mathcal{S}_C$.

The situation is almost identical for R. A strategy $\sigma_R$ specifies the next robber move at time $t$, depending on information available to R at $t$; this information is the realizations $x_0, x_1, \ldots, x_t$ and $y_0, y_1, \ldots, y_{t-1}$. We define the following.

(1) $A_R = V$ is the set of possible R actions (possible robber positions).
(2) $H_R^{(m)} \subseteq V^m \times V^{m-1}$ is the set of *feasible $m$ long sequences* of cop/robber configurations.
(3) $H_R = \bigcup_{m=0}^{\infty} H_R^{(m)}$ is the set of all *finite-length* feasible cop/robber histories.

Hence a robber strategy is a function $\sigma_R : H_R \to \mathbf{P}(A_R)$, i.e., a function which maps to every finite-length history a probability on the next R move; again, we are interested in *feasible* strategies, i.e., those which assign positive probabilities only to feasible next-step cop configurations. We will denote the set of all feasible R strategies by $\mathcal{S}_R$.

A specific strategy pair $(\sigma_R, \sigma_C)$, specifies the probabilities $p(x_0, x_1, \ldots, x_t, y_0, y_1, \ldots y_t \mid \sigma_R, \sigma_C)$ for all cylindrical sets $(X_0 = x_0, X_1 = x_1, \ldots, X_t = x_t, Y_0 = y_0, Y_1 = y_1, \ldots, Y_t = y_t)$. Hence, letting $\widetilde{H}_R = V^{\mathbb{N} \times \mathbb{N}}$ denote the set of all infinitely long feasible cop/robber histories, $(\sigma_R, \sigma_C)$ induces a probability measure on the associated $\sigma$-algebra. In short, *at the start* of the game $\Gamma$,

C chooses $\sigma_C$ and R chooses $\sigma_R$, resulting in a well defined *expected capture time, conditioned on* $\sigma_R$ *and* $\sigma_C$; this quantity is denoted by $E(T \mid \sigma_R, \sigma_C, \Gamma)$.

The strategies $\sigma_R$ and $\sigma_C$ can be used in any truncated game $\Gamma_m$ as well: C and R will use them to generate moves only until turn $m$. Hence the corresponding expected capture time for $\Gamma_m$ is also well defined; it will be denoted by $E(T \mid \sigma_R, \sigma_C, \Gamma_m)$. Clearly $\Gamma$ and every $\Gamma_m$ are two-person, zero-sum games.

It is worth emphasizing that C and R choose their strategies *simultaneously*, before the game starts. Even though the game rules stipulate that (in every turn) C plays before R, this is unimportant because (the probabilities of) both players' moves are specified by their strategies, which have been selected before the game starts. (It is a well known fact [4] that, if either player chooses his optimal strategy then he has no incentive to change it, no matter what strategy the other player uses.) In fact, the rules can be modified so that the players play simultaneously: at the initial turn ($t = 0$) R plays a "null" move and C plays $X_0$; at all subsequent turns ($t \in \mathbb{N}$), R plays $Y_{t-1}$ and C plays $X_t$. The game remains the same under these modified rules.

In $\Gamma$, if R uses $\sigma_R$ and C uses $\sigma_C$ then C pays R $E(T \mid \sigma_R, \sigma_C, \Gamma)$ per game (on the average). The situation is similar in $\Gamma_m$ (for every $m \in \mathbb{N}_0$) except that the game lasts at most $m$ steps and if the robber has not been caught by the end of the $m$-th step, he receives a payoff of $m$; the average payoff in $\Gamma_m$ is $E(T \mid \sigma_R, \sigma_C, \Gamma_m)$. In either case (full or truncated game) R tries to maximize the payoff while C tries to minimize it.

Consider for a moment *pure* strategies available to C in the finite-length game $\Gamma_m$. Each such strategy, call it $s_C$, will be a collection of *deterministic* functions mapping finite length cop histories $x_0, x_1, \ldots, x_{t-1}$ to feasible moves:

$$X_t = f(x_0, x_1, \ldots x_{t-1}).$$

Since $t \le m$ and $|V|$ is finite, there is a finite number of cop histories and a finite number of pure strategies that C can use. Similarly R has a finite number of pure strategies. Hence $\Gamma_m$ is a finite game and has a *value* [4], which however is generally achieved by *mixed* strategies $\widehat{\sigma}_R^{(m)}$ and $\widehat{\sigma}_C^{(m)}$:

$$\max_{\sigma_R \in \mathcal{S}_R^m} \min_{\sigma_C \in \mathcal{S}_C^m} E(T \mid \sigma_R, \sigma_C, \Gamma_m) = E\left(T \mid \widehat{\sigma}_R^{(m)}, \widehat{\sigma}_C^{(m)}, \Gamma_m\right) = \min_{\sigma_C \in \mathcal{S}_C^m} \max_{\sigma_R \in \mathcal{S}_R^m} E(T \mid \sigma_R, \sigma_C, \Gamma_m) \tag{13}$$

(In fact $\Gamma_m$ can be summarized by a finite payoff matrix $Q^{(m)}$ where $Q_{s_R, s_C}^{(m)} = E(T \mid s_R, s_C, \Gamma_m)$ and $s_R, s_C$ are understood as *classes* of pure strategies which are identical up to the $m$-th turn.) We sometimes will denote $E\left(T \mid \widehat{\sigma}_R^{(m)}, \widehat{\sigma}_C^{(m)}, \Gamma_m\right)$ by $val(\Gamma_m)$, for brevity.

In the infinite-length game $\Gamma$ there is an infinite number of pure strategies, hence the existence of a value and optimizing strategies is not guaranteed. Let us define

$$\underline{val}(\Gamma) = \sup_{\sigma_R \in \mathcal{S}_R} \inf_{\sigma_C \in \mathcal{S}_C} E(T \mid \sigma_R, \sigma_C, \Gamma)$$

$$\overline{val}(\Gamma) = \inf_{\sigma_C \in \mathcal{S}_C} \sup_{\sigma_R \in \mathcal{S}_R} E(T \mid \sigma_R, \sigma_C, \Gamma).$$

Hence R, playing optimally, is guaranteed to receive no less than (arbitrarily close to) $\underline{val}(\Gamma)$; C, playing optimally, is guaranteed to pay no more than (arbitrarily close to) $\overline{val}(\Gamma)$; we have $\underline{val}(\Gamma) \le \overline{val}(\Gamma)$. What we want is to show $\underline{val}(\Gamma) = \overline{val}(\Gamma)$; if equality holds, then we will denote the common value by $val(\Gamma)$, the *value* of the game $\Gamma$.

Existence of $val\,(\Gamma)$ follows *almost* immediately from a theorem proved by Gurevich [20]. A couple of modifications of his proof are required. First, we need the following lemma.

**Lemma 4.1.** *Given any graph $G$, let $\overline{\sigma}_C$ be the strategy according to which the cop performs a random walk on $G$. Define $\overline{v} = \sup_{\sigma_R \in \mathcal{S}_R} E\,(T \mid \sigma_R, \overline{\sigma}_C, \Gamma)$. Then*

$$\lim_{m \to \infty} val\,(\Gamma_m) \leq \underline{val}\,(\Gamma) \leq \overline{v} < \infty.$$

*Proof.* From Theorem 2.1 we know that, when the cop random-walks in $G$, he can catch the robber in finite expected time. Hence

$$\sup_{\sigma_R \in \mathcal{S}_R} E\,(T \mid \sigma_R, \overline{\sigma}_C, \Gamma) = \overline{v} < \infty. \tag{14}$$

Then we have

$$\underline{val}\,(\Gamma) = \sup_{\sigma_R \in \mathcal{S}_R} \inf_{\sigma_C \in \mathcal{S}_C} E\,(T \mid \sigma_R, \sigma_C, \Gamma) \leq \sup_{\sigma_R \in \mathcal{S}_R} E\,(T \mid \sigma_R, \overline{\sigma}_C, \Gamma) = \overline{v} < \infty.$$

If we extend the truncated game $\Gamma_m$ by one move we get the game $\Gamma_{m+1}$ and R's payoff cannot decrease. Hence $val\,(\Gamma_m) \leq val\,(\Gamma_{m+1})$, i.e. the sequence $\{val\,(\Gamma_m)\}_{m=0}^{\infty}$ is nondecreasing. It is also bounded, because in the full game $\Gamma$, R can do at least as well as in any truncated game $\Gamma_m$. Hence $val\,(\Gamma_m) \leq val\,(\Gamma_{m+1}) \leq \underline{val}\,(\Gamma)$, from which follows $v = \lim_{m \to \infty} val\,(\Gamma_m) \leq \underline{val}\,(\Gamma)$.   $\square$

Using Lemma 4.1 we can now prove the following.

**Theorem 4.2.** *Given any graph $G$ and the corresponding CiR game $\Gamma$ played with $c\,(G)$ cops, $val\,(\Gamma)$ exists and satisfies*

$$val\,(\Gamma) = \lim_{m \to \infty} val\,(\Gamma_m).$$

*Furthermore, there exists a strategy $\widehat{\sigma}_C$ such that*

$$\sup_{\sigma_R \in \mathcal{S}_R} E\,(T \mid \sigma_R, \widehat{\sigma}_C, \Gamma) = val\,(\Gamma) \tag{15}$$

*and, for every $\varepsilon > 0$ there exists an $m_\varepsilon$ and a strategy $\widehat{\sigma}_R^{\varepsilon}$ such that*

$$\forall m \geq m_\varepsilon : val\,(\Gamma) - \varepsilon \leq \inf_{\sigma_C \in \mathcal{S}_C} E\,(T \mid \widehat{\sigma}_R^{\varepsilon}, \sigma_C, \Gamma_m). \tag{16}$$

*Proof.* The theorem is a rephrasing of Gurevich's Theorem 1 [20] and his proof can also be used here, except for the following two points.

 (1) Gurevich assumes $E\,(T \mid \sigma_R, \sigma_C, \Gamma)$ is bounded:

$$\exists M : \forall\,(\sigma_R, \sigma_C) : |E\,(T \mid \sigma_R, \sigma_C, \Gamma)| \leq M$$

 This is not necessarily true for CiR. But, as he points out [20, p.372], this assumption can be removed, provided the sequence $\{val\,(\Gamma_m)\}_{m=0}^{\infty}$ is bounded; in the CiR game this is true because of Theorem 2.1 (as discussed in the proof of Lemma 4.1).
 (2) Gurevich studies games in which the two players move simultaneously. In CiR this is not the case but, since C never sees R's move, we can (as already mentioned) rearrange the order of moves and have the moves $(Y_{t-1}, X_t)$ take place simultaneously.

Other than these two points, Gurevich's proof applies exactly to the current theorem.   $\square$

**Remark**. From (15) we see that C has an optimal strategy. From (16) we see that, for every $\varepsilon > 0$, R has an "$\varepsilon$-optimal" strategy, which is *uniformly* good (i.e., for *all* $m \geq m_\varepsilon$).

We now can replace equation (3) with a rigorous definition of $ct_i\,(G)$.

**Definition 4.3.** *Given a graph $G$, we define the* invisible capture time *of $G$ to be*

$$ct_i(G) = val(\Gamma)$$

*where the game $\Gamma$ is played on $G$ with $c(G)$ cops.*

Hence $ct_i(G)$ is not necessarily an *achieved* expected capture time, but it can be approximated within any $\varepsilon > 0$ by using strategies $\widehat{\sigma}_R^\varepsilon$ and $\widehat{\sigma}_C$.

4.2. **Drunk Robber.** We now turn to the drunk variant of CiR. Our goal is to rigorously define $dct_i(G)$. Since a single player is involved, the situation is simpler than in the adversarial variant. In fact, as we will now explain, the drunk variant of CiR is a *partially observed Markov decision process* (POMDP) [31, 35, 48].

The robber process $Y_t$ is a Markov chain on $V$, with transition probability matrix $P$. As explained in [26], the joint process $(X_t, Y_t)$ is also a Markov process on the extended state space $(V \times V) \cup \{\lambda\}$, where $\lambda$ is the *capture state*. $(X_t, Y_t)$ is governed by the *controlled* transition probability matrix $P(U_t)$, where $U_t = X_t$ is the *control* variable (selected by C) which changes the transition probabilities [26]. In particular, $P(U_t)$ assigns probability 1 to transitions from "diagonal" states $(X_t, X_t)$ to the capture state $\lambda$. C's goal is to select the sequence $U_0, U_1, \ldots$ so as to minimize expected capture time. The state $(X_t, Y_t)$ is *partially* observable by C, since he never knows $Y_t$.

We will use the notation of Section 4.1 denoting, however, the full one-player game by $\overline{\Gamma}$ and the truncated one-player game by $\overline{\Gamma}_m$. We also need the following facts.

(1) $\mathcal{S}_C$, the space of cop strategies, is compact. This is so by Tychonoff's theorem, since $\mathcal{S}_C$ is a product of probability spaces and the probabilities are on finite event sets.
(2) For every $m \in \mathbb{N}_0$, $E(T \mid \sigma_C, \overline{\Gamma}_m)$ is a *continuous* function of $\sigma_C$ (with domain $\mathcal{S}_C$). This is the case because $E(T \mid \sigma_C, \overline{\Gamma}_m)$ depends continuously on a *finite* number of variables.

C must select a strategy $\sigma_C$ which minimizes

$$E\left(T \mid \sigma_C, \overline{\Gamma}\right) = E\left(\sum_{t=0}^{\infty} \mathbf{1}\left(X_t \neq Y_t\right) \mid \sigma_C, \overline{\Gamma}\right);$$

here $\mathbf{1}(X_t \neq Y_t)$ is the indicator function of the event $X_t \neq Y_t$; this is a typical *infinite horizon, undiscounted* POMDP problem. Such problems have been studied by several authors [43, 47] who prove the existence of a minimizing strategy for a quite general setup using rather involved proofs. We present a simpler and shorter proof, based on the reduction of Gurevich's argument [20] to the one-player case. The proof is useful to illustrate the issues involved, and may also have some independent interest.

Note that, similarly to the adversarial variant, $\Pr(T \mid \sigma_C, \overline{\Gamma}_m)$ is well defined (for every $\sigma_C$) and can be extended to $\Pr(T \mid \sigma_C, \overline{\Gamma})$. Hence $E(T \mid \sigma_C, \overline{\Gamma}_m)$, $E(T \mid \sigma_C, \overline{\Gamma})$ are well defined for every $\sigma_C \in \mathcal{S}_C$. Let us define

$$val\left(\overline{\Gamma}_m\right) = \inf_{\sigma_C \in \mathcal{S}_C} E\left(T \mid \sigma_C, \overline{\Gamma}_m\right),$$

$$val\left(\overline{\Gamma}\right) = \inf_{\sigma_C \in \mathcal{S}_C} E\left(T \mid \sigma_C, \overline{\Gamma}\right).$$

We then have the following.

**Theorem 4.4.** *Given any graph $G$ and the corresponding CiR game $\Gamma$, we have*

$$val\left(\overline{\Gamma}\right) = \lim_{m \to \infty} val\left(\overline{\Gamma}_m\right).$$

*Furthermore, there exists a strategy $\widetilde{\sigma}_C$ such that*

$$E\left(T \mid \widetilde{\sigma}_C, \overline{\Gamma}\right) = val\left(\overline{\Gamma}\right).$$

*Proof.* We have

$$\inf_{\sigma_C \in \mathcal{S}_C} E\left(T \mid \sigma_C, \overline{\Gamma}_m\right) \leq \inf_{\sigma_C \in \mathcal{S}_C} E\left(T \mid \sigma_C, \overline{\Gamma}_{m+1}\right) \leq \inf_{\sigma_C \in \mathcal{S}_C} E\left(T \mid \sigma_C, \overline{\Gamma}\right).$$

Hence

$$val\left(\overline{\Gamma}_m\right) \leq val\left(\overline{\Gamma}_{m+1}\right) \leq val\left(\overline{\Gamma}\right) \leq E\left(T \mid \overline{\sigma}_C, \overline{\Gamma}\right) < \infty,$$

where $\overline{\sigma}_C$ is the random-walking strategy, which is guaranteed to capture the robber in finite expected time. Hence $v = \lim_{m \to \infty} val\left(\overline{\Gamma}_m\right)$ exists and $v \leq val\left(\overline{\Gamma}\right)$.

Now, in $val\left(\overline{\Gamma}_m\right) = \inf_{\sigma_C} E\left(T \mid \sigma_C, \overline{\Gamma}_m\right)$, the infimum is achieved by some $\widetilde{\sigma}_C^{(m)}$ (since $E\left(T \mid \sigma_C, \overline{\Gamma}_m\right)$ is a continuous function of the $\sigma_C$ probabilities, which take values in a compact set). Define

$$K_m = \left\{\sigma_C : E\left(T \mid \sigma_C, \overline{\Gamma}_m\right) \leq v\right\}.$$

Since $E\left(T \mid \widetilde{\sigma}_C^{(m)}, \overline{\Gamma}_m\right) \leq v$, every $K_m$ is nonempty. Also, $val\left(\overline{\Gamma}_m\right) \leq val\left(\overline{\Gamma}_{m+1}\right) \leq v$ implies that $K_{m+1} \subseteq K_m$. As mentioned, $E\left(T \mid \sigma_C, \overline{\Gamma}_m\right)$ is a continuous function on $\mathcal{S}_C$ and $\mathcal{S}_C$ is compact. For every $m$, $K_m$ is the preimage of the compact set $[0, v]$, hence $K_m$ is compact. It follows that

$$K_\infty = \bigcap_{m=0}^{\infty} K_m$$

is nonempty. So let us take some $\widetilde{\sigma}_C \in K_\infty$. Then

$$
\begin{aligned}
E\left(T \mid \widetilde{\sigma}_C, \overline{\Gamma}\right) &= \lim_{m \to \infty} \sum_{t=0}^{m} t \cdot \Pr\left(T = t \mid \widetilde{\sigma}_C, \overline{\Gamma}\right) \\
&= \lim_{m \to \infty} \sum_{t=0}^{m} t \cdot \Pr\left(T = t \mid \widetilde{\sigma}_C, \overline{\Gamma}_m\right) = \lim_{m \to \infty} E\left(T \mid \widetilde{\sigma}_C, \overline{\Gamma}_m\right) \leq v.
\end{aligned}
$$

In other words, we have $val\left(\overline{\Gamma}\right) \leq v$. Hence $val\left(\overline{\Gamma}\right) = v$ and is achieved by $\widetilde{\sigma}_C$. $\qquad\square$

We now provide a formal definition of $dct_i\left(G\right)$ (to replace equation (4)).

**Definition 4.5.** *Given a graph $G$, we define the* invisible drunk capture time *of $G$ to be*

$$dct_i\left(G\right) = val\left(\overline{\Gamma}\right)$$

*where the game $\overline{\Gamma}$ is played on $G$.*

4.3. **Discussion.** The results of Section 4.2 show that for every graph $G$ there is an optimal average capture time $dct_i(G)$ and the cop player can catch the drunk robber in $dct_i(G)$ turns of the game (on the average) if he plays the optimal strategy $\widetilde{\sigma}_C$. The actual *computation* of $dct_i(G)$ and $\widetilde{\sigma}_C$ is the subject of ongoing research in the POMDP community. While the problem is well understood in principle (an *optimality equation* must be solved, similar to the one presented in [26] for the CR problem) and despite much effort, currently available POMDP algorithms are not computationally viable, even for moderate-size graphs (the interested reader is referred to [32, 35] for an extensive discussion of the issues involved).

The results of Section 4.1 yield similar conclusions for the adversarial robber, with one difference. Namely, while an optimal strategy $\widehat{\sigma}_C$ is available to C, the best R can do is to approximate $ct_i(G)$ within $\varepsilon$ (for any $\varepsilon > 0$) by using an $\varepsilon$-optimal strategy $\widehat{\sigma}_R^\varepsilon$. In the computational direction, even less progress has been achieved than in the drunk robber case (see [45]).

We consider the development of efficient CiR algorithms important, especially from the applications point of view. Further, since exact algorithms quickly become intractable (even for relatively small graphs), we believe that the solution will be obtained by algorithms which are approximate but have performance guarantees. Such algorithms will probably make use of domain-specific heuristics.

## 5. Some Bounds for General Graphs

In this section we provide a lower bound on $dct_i(G)$ and an upper bound on $ct_i(G)$; both bounds hold for any $G$. These results have independent interest and will also be used in Section 6 to obtain $dct_i(G)$ and $ct_i(G)$ for special graph families. As already stated, we assume that $K$, the number of cops, equals $c(G)$.

### 5.1. **General lower bound of $dct_i(G)$.**
Let $G = (V, E)$ be any graph and $(X_t)_{t \in \mathbb{N}_0}$ any searching schedule for $K \in \mathbb{N}$ cops (the $k$th cop, $k \in [K]$, moves from $X_{t-1}^k$ to $X_t^k$ at time $t \in \mathbb{N}$). For $t \in \mathbb{N}_0$, let $Z_t = \bigcup_{k \in [K]} \{X_t^k\}$ be the set of vertices occupied by the cops at the end of turn $t$.

We now define certain conditional probabilities which will be used to obtain the lower bound on $dct_i(G)$.

$\bar{p}_t(v)$ : Pr("at $t$-th turn, after the cop move, robber is at $v$" | "robber has not been captured"),

$\hat{p}_t(v)$ : Pr("at $t$-th turn, after the robber move, robber is at $v$" | "before a possible capture"),

$p_t(v)$ : Pr("at the completion of $t$-th turn, robber is at $v$" | "robber has not been captured").

The following remarks should make clearer the meaning of the above probabilities. Effectively, we break each turn of the game into three phases.

(1) In the first phase, the cops move and they may capture the robber or not. The probability that the robber is at $v$, *given that he has not been captured*, is $\bar{p}_t(v)$.

(2) In the second phase, the robber moves but a possible capture (i.e., if he entered a vertex occupied by a cop) is not yet effected. Hence $\hat{p}_t(v)$ is the probability that the robber is at $v$ (*even if $v$ contains a cop*).

(3) In the final phase possible captures (i.e., if the robber ran into a cop) are effected and $p_t(v)$ is the probability that the robber is in $v$ *given that no capture took place.*

So finally, the probability that the robber is caught at time $t \in \mathbb{N}$ is

$$c_t = \sum_{u \in Z_t} \Big( p_{t-1}(u) + \hat{p}_t(u) \Big).$$

We will now write the equations which govern the evolution of $\bar{p}_t(v)$, $\hat{p}_t(v)$, $p_t(v)$. It is easy to see that, for every $v \in V$, $\bar{p}_0(v) = 0$ (the robber has not entered the graph yet). Note that $\bar{p}_0$ is not a probability distribution; one can think of it as a useful function to get the desired recursion started. It is easy to see that $\hat{p}_t(v) = 1/n$. Regarding $p_0(v)$ we have $p_0(v) = 0$ for $v \in Z_0$, and

$$p_0(v) = \frac{\hat{p}_0(v)}{1 - \sum_{u \in Z_0} \hat{p}_0(u)} = \frac{1/n}{1 - |Z_0|/n} = \frac{1}{n - |Z_0|}$$

for $v \in V \setminus Z_0$.

Suppose now that, at a given time $t \in \mathbb{N}$, the game is still on; that is, the robber is still hiding somewhere on the graph $G$. Suppose that we know the distribution of the position of the drunk robber, $p_{t-1} : V \to [0,1]$, at the end of the previous turn (let us repeat that this probability is *conditional on the robber not having been captured*). The cops move from $X_{t-1}$ to $X_t$, and so they capture the robber with probability $\sum_{u \in Z_t} p_{t-1}(u)$. Conditioning on the fact that the robber is still not captured, let $\bar{p}_t(v)$ be the probability that the robber is at vertex $v$ after this cop move. We have $\bar{p}_t(v) = 0$ for $v \in Z_t$, and

$$\bar{p}_t(v) = \frac{p_{t-1}(v)}{1 - \sum_{u \in Z_t} p_{t-1}(u)} \tag{17}$$

for $v \in V \setminus Z_t$.

Now, the robber performs a step of his random walk. Let $\hat{p}_t(v)$ be the probability that he is at vertex $v$ after this move; that is,

$$\hat{p}_t(v) = \sum_{u \in N(v)} \frac{\bar{p}_t(u)}{\deg(u)}. \tag{18}$$

The probability of the robber being captured at the completion of his move is $\sum_{u \in Z_t} \hat{p}_t(u)$. Assuming, as before, that the robber is still lucky at the end of turn $t$, we get the formula for the distribution of the robber's position at the end of turn $t$ (once again, conditioned on the robber not having been captured). For $v \in Z_t$: $p_t(v) = 0$; for $v \in V \setminus Z_t$:

$$p_t(v) = \frac{\hat{p}_t(v)}{1 - \sum_{u \in Z_t} \hat{p}_t(u)}. \tag{19}$$

Eqs.(17)-(19) along with the initial conditions for $t = 0$, describe the evolution of $\bar{p}_t(v)$, $\hat{p}_t(v)$, $p_t(v)$ for a given strategy of the cops. Now, we are ready to state a lower bound for $dct_i(G)$ for a general graph $G$.

**Lemma 5.1.** *Let $G$ be any graph on $n > c(G)$ vertices with maximum degree $\Delta = \Delta(G)$ and minimum degree $\delta = \delta(G)$ such that $\frac{\Delta K}{\delta(n-K)} \le 1/24$. Then,*

$$dct_i(G) \ge \frac{\delta \cdot (n - K)}{7e\Delta \cdot c(G)}.$$

*Proof.* We will use the notation introduced earlier in this subsection. Let $G = (V, E)$ be any graph, and suppose that $K = c(G)$ cops try to catch the drunk robber on this graph. We can assume that $n > K$; the statement is trivially true otherwise. Let us define the following deterministic function: $M_0 = \frac{\Delta/\delta}{n-K}$, and for $t \in \mathbb{N}$

$$M_t = \frac{M_{t-1}}{1 - 2KM_{t-1}}.$$

We will show that for any $t \in \mathbb{N}_0$ and any vertex $v \in V$ we have

$$\max\left\{\bar{p}_t(v), \hat{p}_t(v), p_t(v)\right\} \le M_t \frac{\deg(v)}{\Delta}. \tag{20}$$

We prove the claim by induction. Since for each $v \in V$ we have

$$\max\left\{\bar{p}_0(v), \hat{p}_0(v), p_0(v)\right\} \le \frac{1}{n - K} = M_0 \frac{\delta}{\Delta} \le M_0 \frac{\deg(v)}{\Delta},$$

the base case ($t = 0$) holds. Suppose now that (20) holds for $t - 1 \in \mathbb{N}_0$. It follows immediately from (17) that for any $v \in V$

$$\bar{p}_t(v) \leq \frac{p_{t-1}(v)}{1 - \sum_{u \in Z_t} p_{t-1}(u)} \leq \frac{M_{t-1} \frac{\deg(v)}{\Delta}}{1 - K M_{t-1}} \leq M_t \frac{\deg(v)}{\Delta}.$$

From (18) we get that

$$
\begin{aligned}
\hat{p}_t(v) &= \sum_{u \in N(v)} \frac{\bar{p}_t(u)}{\deg(u)} \leq \sum_{u \in N(v)} \frac{M_{t-1} \frac{\deg(u)}{\Delta}}{\deg(u)(1 - K M_{t-1})} \\
&= \sum_{u \in N(v)} \frac{M_{t-1}}{\Delta(1 - K M_{t-1})} = \frac{M_{t-1} \frac{\deg(v)}{\Delta}}{1 - K M_{t-1}} \leq M_t \frac{\deg(v)}{\Delta}
\end{aligned}
$$

for any $v \in V$, so the very same bound holds for $\hat{p}_t(v)$. Finally, from (19) we get that

$$p_t(v) \leq \frac{\hat{p}_t(v)}{1 - \sum_{u \in Z_t} \hat{p}_t(u)} \leq \frac{M_{t-1} \frac{\deg(v)}{\Delta}}{1 - K M_{t-1}} \left( 1 - \frac{K M_{t-1}}{1 - K M_{t-1}} \right)^{-1} = \frac{M_{t-1} \frac{\deg(v)}{\Delta}}{1 - 2 K M_{t-1}} = M_t \frac{\deg(v)}{\Delta},$$

and the proof of the claim is complete.

Suppose that $M_{t-1} < \frac{3\Delta}{\delta(n-K)}$ (which implies that $M_s < \frac{3\Delta}{\delta(n-K)}$ for $0 \leq s < t$, since $M_s$ is increasing with $s$). Then,

$$M_t \leq M_{t-1} \left( 1 - \frac{6\Delta K}{\delta(n-K)} \right)^{-1} \leq \ldots \leq M_0 \left( 1 - \frac{6\Delta K}{\delta(n-K)} \right)^{-t} \leq \frac{\Delta}{\delta(n-K)} \exp \left( \frac{7\Delta K t}{\delta \cdot (n-K)} \right),$$

where the last inequality follows since $\frac{1}{1-x} \leq e^{7x/6}$ for $x \leq 1/4$ and $\frac{\Delta K}{\delta(n-K)} \leq 1/24$. Therefore, it takes at least $\tau = \frac{\delta(n-K)}{7\Delta K}$ steps for $M_t$ to reach $\frac{3\Delta}{\delta(n-K)}$. During this time period, the probability that we catch the robber at a given time $t \leq \tau$ is

$$c_t = \sum_{u \in Z_t} \left( p_{t-1}(u) + \hat{p}_t(u) \right) \leq 2 K M_t \leq \frac{6 K \Delta}{\delta(n-K)}.$$

Hence, the probability that the robber is still not caught at time $\tau$ is at least

$$\left( 1 - \frac{6 K \Delta}{\delta(n-K)} \right)^{\tau} \geq e^{-1}.$$

Finally, we get that the expected capture time is at least $\tau/e$, and the proof is finished.     $\square$

### 5.2. General upper bound of $ct_i(G)$.
In this subsection, we investigate the adversarial robber case providing a universal upper bound for the capture time.

**Lemma 5.2.** *Let $G$ be any graph of $n$ vertices with maximum degree $\Delta = \Delta(G)$, cop number $c(G)$, and diameter $D = D(G)$. Let $\hat{T}$ denote the maximum number of steps it takes for $c(G)$ cops to catch the visible robber, when they use an optimal strategy (independently of the position of the robber). Then,*

$$ct_i(G) \leq (\hat{T} + D) \cdot (\Delta + 1)^{\hat{T}} \cdot n. \tag{21}$$

*Proof.* We will introduce a cop strategy by which the adversarial robber will be captured in at most $(\widehat{T} + D) \cdot (\Delta + 1)^{\widehat{T}} \cdot n$ time steps (even if he knows the strategy in advance and plays optimally against it).

The cop strategy is executed in *rounds*, each round consisting of one or more steps of the game. In each round the cops first move to those vertices from which they can capture the visible robber (independently of his position) in at most $\widehat{T}$ steps. (Such a $\widehat{T} < \infty$ exists for every $G$, since $c(G)$ cops suffice to capture the visible robber in finite time.) Note that in the first round the cops start immediately from the aforementioned vertices. At any rate, once the cops are there, they uniformly at random guess the current position of the robber, and then at each of the next $\widehat{T}$ steps they uniformly at random guess the behaviour of the robber and apply their optimal move for this guessed behaviour. The round is over $\widehat{T}$ steps after the cops arrived at their optimal starting position. If the robber has not been captured by then, the cops start the next round, following the same strategy.

In each round, it takes at most $D$ steps for the cops to go back to the optimal starting position; after that they move $\widehat{T}$ additional steps in which the robber might be caught. In order to catch the robber during one round, it suffices that the cops guess correctly the position of the robber at the start of the round (which happens with probability $\frac{1}{n}$) and also guess correctly the move of the robber in each of the following $\widehat{T}$ steps. Since the robber at each vertex has at most $\Delta + 1$ choices (he can move to any of the at most $\Delta$ neighbours, or he can stand still), in each step the probability of the cops making the right choice is $\frac{1}{\Delta+1}$. Thus, capture in a round happens with probability at least $\frac{1}{n \cdot (\Delta+1)^{\widehat{T}}}$. Since the bounds on the number of a round hold independently of the starting point, consecutive rounds are independent, and the expected number of rounds until the robber is caught is at most $n \cdot (\Delta + 1)^{\widehat{T}}$. Hence, $ct_i \leq (\widehat{T} + D) \cdot (\Delta + 1)^{\widehat{T}} \cdot n$ and the proof is complete. $\qquad\square$

**Remark.** The bound of (21) does *not* mean that $ct_i(G)$ is $O(n)$. Both $\Delta$ and $\widehat{T}$ depend on $G$ and hence on $n$, the number of vertices.

## 6. The Cost of Drunkenness for Special Graph Families

We will now show that, by restricting ourselves to some particular graph families, we can obtain non-trivial bounds on the iCOD. Paths and cycles were considered in [26], so we state the results only (Subsection 6.1). The complete $d$-ary tree of depth $L$ is studied next (Subsection 6.2), and then we study the grid (Subsection 6.3). Finally, in Subsection 6.4, we give an example of a family of graphs (brooms) showing that $F_i(G)$ can be arbitrarily close to any value in $[2, \infty)$.

### 6.1. **Paths and Cycles.** The following two theorems have been proved in [26].

**Theorem 6.1.** *Let $P_n$ be the path of $n$ vertices. Then $ct_i(P_n) = n - 1$ and*

$$\frac{n}{2}\left(1 - O\left(\frac{\log n}{n}\right)\right) \ \leq \ dct_i(P_n) \ \leq \ \frac{n-1}{2}.$$

*In particular, $dct_i = (1 + o(1)) \cdot \frac{n}{2}$ and the cost of drunkenness is*

$$F_i(G) = \frac{ct_i(P_n)}{dct_i(P_n)} = 2 + o(1).$$

**Theorem 6.2.** *Let $C_n$ be the cycle of $n$ vertices. Then $ct_i(C_n) = \frac{n-1}{2}$ and*

$$\frac{n}{4}\left(1 - O\left(\frac{\log n}{n}\right)\right) \leq dct_i(C_n) \leq \frac{n-1}{4}.$$

*In particular, $dct_i(C_n) = (1 + o(1))\frac{n}{4}$ and the cost of drunkenness is*

$$F_i(G) = \frac{ct_i(C_n)}{dct_i(C_n)} = 2 + o(1).$$

6.2. **Trees.** We restrict ourselves to $T_{d,L}$, the complete $d$-ary tree of depth $L$, for $d \geq 2$. This tree has $n = \frac{d^{L+1}-1}{d-1}$ vertices and $M = d^L$ leaves. We know that one cop suffices to capture the robber on any tree, so let us consider a game played by a single cop and a robber. In this subsection, we will show the following results.

**Theorem 6.3.** *Let $G = T_{d,L}$ be the complete $d$-ary tree of depth $L$ with $n = \frac{d^{L+1}-1}{d-1}$ vertices. Then,*

$$2Ld^{L-1} \cdot (1 - o(1))\frac{d-1}{d} \leq ct_i(G) \leq 2Ld^{L-1} \cdot (1 - o(1)).$$

*In particular, $ct_i(G) = \Theta(n \log n)$.*

**Theorem 6.4.** *Let $G = T_{d,L}$ be the complete $d$-ary tree of depth $L$ with $n = \frac{d^{L+1}-1}{d-1}$ vertices. Then,*

$$dct_i(G) = \Theta(n).$$

The following corollary is an immediate implication of these two theorems.

**Corollary 6.5.** *Let $G = T_{d,L}$ be the complete $d$-ary tree of depth $L$ with $n = \frac{d^{L+1}-1}{d-1}$ vertices. The cost of drunkenness of $G$ is*

$$F_i(G) = \frac{ct_i(G)}{dct_i(G)} = \Theta(\log n).$$

*Proof of Theorem 6.3.* Since $c(G) = 1$, the (invisible) robber is chased by a single cop. To simplify the notation we use $X_t = X_t^1$ for a position of the cop at time $t$ (this time the vector $X_t$ has one coordinate).

In order to give an upper bound on $ct_i(G)$, we provide a strategy for the cop and show that, independently of the robber's behaviour, in expectation, the cop will catch the robber after at most a certain number of steps. Denote by a *preleaf* a vertex at distance 1 from any leaf. Consider the following strategy of the cop:

(1) Start at the root,
(2) Choose uniformly at random a preleaf $v$ and go there,
(3) Choose a random permutation of the leaves below $v$ and visit them in this order, always returning to the preleaf $v$,
(4) Return to the root,
(5) Repeat from step 2 down.

Similarly to Lemma 5.2, a *round* is a sequence of steps starting at the root, visiting a preleaf and all its leaves and going back to the root. Note that each round consists of $2L + 2(d-1)$ steps. Observe that the cop's strategy in step (2) is equivalent to choosing at each layer $0 \leq i \leq L-2$ a random vertex among all neighbours at layer $i+1$ with probability $\frac{1}{d}$. Note that, independently of the robber's strategy, he is caught in each round with probability $\frac{1}{d^{L-1}}$. Indeed, provided that the robber is in the subtree below the cop (at some step $0 \leq i \leq L-2$), the probability that this is also true at the next round is $1/d$. If this property is preserved from the beginning of the round until step $L-2$, the robber has no chance to survive and is caught after visiting all leaves.

Since all rounds are independent, the expected number of rounds needed to capture the robber using this strategy is $d^{L-1}$, and thus

$$ct_i \leq d^{L-1}(2L + 2(d-1)) - L - (d-1), \tag{22}$$

since in the last round the cop does not have to return to the root (and so he saves $L$ moves), and he has to visit only half of the leaves (in expectation) before catching the robber (and so he saves another $2(d-1) - 2\frac{d}{2} = d-1$ moves, in expectation).

For the lower bound, we provide a strategy for the robber against which any cop will need at least a certain number of rounds, in expectation. The robber strategy depends on the cop's moves and can be briefly described as follows: the robber always tries to be (after his move) at distance 2 from the cop. As a result, the distance between players is never larger than 3. Moreover, the robber is trying to stay at the layer above the cop, if it is possible. Let us now describe the robber strategy in more detail.

(1) **For $t = 0$.** After the cop decides to start at $X_0$, the robber selects a vertex $Y_0$ at distance 2 from the cop.
   (a) If $X_0$ is located at layer $i \geq 2$, then $Y_0$ is the unique vertex at the layer $i-2$.
   (b) If $X_0$ is at the first layer, then the robber chooses (uniformly at random) a vertex that is also at the first layer but is different from $X_0$.
   (c) Finally, if $X_0$ is the root of $T_{d,L}$, then the robber chooses (uniformly at random) any vertex at the second layer.
(2) **For $t \geq 1$.** The cop moves from $X_{t-1}$ to $X_t$, the robber is at $Y_{t-1}$, and is about to move. There are a number of possibilities to deal with.
   (a) $dist(X_t, Y_{t-1}) = 0$: the game ends.
   (b) $dist(X_t, Y_{t-1}) = 1$ and no neighbour of $Y_{t-1}$ is at distance 2 from $X_t$ (the robber is at a leaf): the robber stays at the same vertex; that is, $Y_t = Y_{t-1}$.
   (c) $dist(X_t, Y_{t-1}) = 1$ and there is a neighbour of $Y_{t-1}$ at distance 2 from $X_t$: the robber chooses (uniformly at random) a vertex at distance 2 from $X_t$ that is as close to the root as possible.
   (d) $dist(X_t, Y_{t-1}) = 2$: the robber does nothing; that is, $Y_t = Y_{t-1}$.
   (e) $dist(X_t, Y_{t-1}) = 3$: there is a unique neighbour of $Y_{t-1}$ at distance 2 from $X_t$; the robber goes there.

It is easy to check that, under this strategy, the distance between the cop and the robber never becomes greater than 3.

Assume that the robber uses the randomized strategy described above; we will show that, even if the cop is aware of this, the capture time will be $\Omega(n \log n)$ and so $ct_i(G)$ will be of at least the same order. Observe that the the robber will only be caught in a leaf. Moreover, any optimal cop strategy must start either at the root or at a neighbour of the root, as otherwise the robber chooses a vertex above the cop and the cop is forced to move towards the root. If the cop starts at the root, then he catches the robber on a leaf if, when in layers $0 \leq i \leq L-2$, he chooses always the subtree of the robber. The advantage of starting at a neighbour of the root is this: since the cop knows the strategy used by the robber, he infers that the robber is in one of the $d-1$ other subtrees (not $d$, as before). This is a much bigger advantage comparing to the additional step back to the root. Hence, this strategy is slightly better and we will consider only this one. (Of course, this applies to the first round of moves only, so starting from the root has exactly the same asymptotic expected capture time.) Note that the probability of choosing the right subtree after going back to the root is $\frac{1}{d-1} \cdot \left(\frac{1}{d}\right)^{L-2}$, since at consecutive layers there is no possibility to obtain partial information about the position of the robber except by checking leaves. The cop does not necessarily have to check all leaves (although this is clearly the best strategy), and thus in order to get a lower bound we will not count the time it takes to check these leaves (we count only the time it takes him to check one leaf). Also, after having exploited all leaves of a preleaf (we can assume the cop did this, since we do not count these extra steps), the cop has to turn back to the root to continue exploiting other subtrees, as otherwise the robber will never be caught.

By the same argument, avoiding the subtree the cop comes from, the probability that in the next sequence of steps from the root to a leaf the cop catches the robber is again $\frac{1}{d-1}\left(\frac{1}{d}\right)^{L-2}$. Since one way to one leaf and back takes at least $2L$ steps, the expected capture time for any cop strategy is at least

$$ct_i(G) \geq 2L \cdot (d-1) \cdot d^{L-2} - L, \tag{23}$$

since in the last round the cop does not have to get back to the root. Thus, combining (22) and (23), we see that $ct_i(G) = \Theta(Ld^L) = \Theta(n \log n)$. $\qquad\square$

*Proof of Theorem 6.4.* We consider the drunken robber performing a random walk on $G = T_{d,L}$, starting from a vertex chosen uniformly at random. The lower bound follows from Lemma 5.1, since $\Delta(G) = d + 1$, $\delta(G) = 1$, and $c(G) = 1$. We have

$$dct_i(G) \geq \frac{n-1}{7e \cdot (d+1)} = \Omega(n). \tag{24}$$

For an upper bound on $dct_i(G)$, we analyze the expected capture time of a cop standing still at the root vertex. Denote by $e_j$ the expected capture time if the robber starts at level $j$ (the root is considered to be at level 0). By definition, $e_0 = 0$, $e_L = 1 + e_{L-1}$ and

$$e_j = 1 + \frac{1}{d+1}e_{j-1} + \frac{d}{d+1}e_{j+1}$$

for $1 \leq j \leq L-1$. Since $e_{L-1} = 1 + \frac{d}{d+1}(1 + e_{L-1}) + \frac{1}{d+1}e_{L-2}$, we have $e_{L-1} = 2d + 1 + e_{L-2}$. Similarly, $e_{L-2} = 2d^2 + 2d + 1 + e_{L-3}$, and in general

$$e_j = 2\sum_{k=0}^{L-j} d^k - 1 + e_{j-1}$$

for $1 \leq j \leq L-1$. Thus,

$$
\begin{aligned}
e_1 &= 2\sum_{k=0}^{L-1} d^k - 1, \\
e_2 &= 2\sum_{k=0}^{L-2} d^k - 1 + 2\sum_{k=0}^{L-1} d^k - 1, \\
e_{L-1} &= 2\sum_{r=1}^{L-1}\sum_{k=0}^{r} d^k - (L-1) = 2\sum_{r=1}^{L-1}\frac{d^{r+1} - 1}{d-1} - L + 1
\end{aligned}
$$

and,

$$
\begin{aligned}
e_L &= 2\sum_{r=1}^{L-1}\frac{d^{r+1}-1}{d-1} - L + 2 = \frac{2}{d-1}\sum_{r=2}^{L}d^r - \frac{2L-2}{d-1} - L + 2 \\
&= \frac{2d^{L+1}-2}{(d-1)^2} - \frac{2+2d}{d-1} - \frac{2L-2}{d-1} - L + 2 \\
&= \frac{2d^{L+1}-2}{(d-1)^2} - \frac{2d+2L}{d-1} - L + 2.
\end{aligned}
$$

Since $e_i \geq e_{i-1}$ for all $1 \leq i \leq L$,

$$
dct_i(G) \leq e_L = O(n). \tag{25}
$$

Combining (24) with (25), we obtain $dct_i(G) = \Theta(n)$, and the proof is complete. $\qquad\square$

6.3. **Grids.** In this subsection, we investigate the square grid $P_N \square P_N$. The result can be easily generalized to rectangle grids. The adversarial robber case seems to be non-trivial. Our upper bound follows from the general upper bound provided in Lemma 5.2. We have no good lower bound so determining the order of $ct_i(P_N \square P_N)$ for a grid remains an open problem. We do much better in the drunk robber case showing that $dct_i(P_N \square P_N)$ is linear.

**Theorem 6.6.** *Let $P_N \square P_N$ be the square $N \times N$-grid with $n = N^2$ vertices. Then*

$$
ct_i(P_N \square P_N) = O(n^{3/2}5^{\sqrt{n}}).
$$

*Proof.* Since $D = 2N$, $\widehat{T} = N - 1$, $\Delta = 4$, it follows from Lemma 5.2 that

$$
ct_i(P_N \square P_N) \leq (3N-1)n5^{N-1} = O(n^{3/2}5^{\sqrt{n}}),
$$

and the result holds. $\qquad\square$

**Theorem 6.7.** *Let $P_N \square P_N$ be the square $N \times N$-grid with $n = N^2$ vertices. Then*

$$
dct_i(P_N \square P_N) = \Theta(n).
$$

*Proof.* Since $c(P_N \square P_N) = 2$, $\Delta = 4$, and $\delta = 2$, it follows from Lemma 5.1 that $dct_i(G) \geq \frac{n}{28e} = \Omega(n)$ and thus the lower bound follows.

On the other hand, for the upper bound, consider the strategy of two cops standing still at antipodal corners of the grid. Assume, without loss of generality, that the first cop is at the top right corner of the grid, and the second one at the bottom left corner. For $i \in \{1,2\}$, denote by $d_i(t) = dist(X_t^i, Y_t)$ the distance of the $i$-th cop from the robber at time $t$. Set $d_i(t) = 0$, if the robber was caught by time $t$ by cop $i$. Note that $d_i(t) \leq 2N$ for any $t$. Since the robber is performing a random walk, on any interior point of the grid and also on the two antipodal corners not occupied by the cops, with probability $\frac{1}{2}$ the value of $d_i(t)$ increases by 1 and $d_{3-i}(t)$ simultaneously decreases by 1 (for both $i \in \{1,2\}$). On the remaining points of the border, the probability that $d_i(t)$ increases by 1 and simultaneously $d_{3-i}$ decreases by 1 is $\frac{2}{3}$ if $d_i(t) < d_{3-i}(t)$, and it is $\frac{1}{3}$ otherwise, for both $i \in \{1,2\}$. Thus, the robber's movements can be seen as moves on the truncated integer line $\{0, 2N\}$, corresponding to the distance of (say) the first cop, and the first time the robber hits any of the two endpoints, he is caught by one cop. Our goal is now to couple the standard random walk (the walk starting at any integer position, and going with probability $\frac{1}{2}$ to the right and with probability $\frac{1}{2}$ to the left) on this truncated integer line with the robber's movements, taking into account the special border conditions of the grid. Let $(Z_t)$ be a sequence of independent random variables, where for each $t$, $\Pr(Z_t = 1) = \Pr(Z_t = -1) = \frac{1}{2}$. Let $(B_t)$ be another sequence of random variables coupled with $Z_t$, taking into account the different transition possibilities on the border: provided that the robber is not yet caught by time $t$, set $B_t = 0$ if $Y_{t-1}$ is either in the inside of the grid or one of the two corner points not occupied by the cops, and otherwise set $B_t = (d_1(t) - d_1(t-1) - Z_t)$. Interpreting a positive value of $Z_t$ as a step of the robber going down or going left and a negative value as a step going up or going right at time $t$, and recalling that $d_1$ increases if and only if $d_2$ decreases, we have

$$
\begin{aligned}
d_1(t) &= d_1(0) + \sum_{j=1}^{t} Z_j + \sum_{j=1}^{t} B_j \\
d_2(t) &= d_2(0) - \sum_{j=1}^{t} Z_j - \sum_{j=1}^{t} B_j,
\end{aligned}
$$

provided that the game is still on. Since $\sum_{j=1}^{t} B_j$ is the same for both values, at any time $t$ either

$$
d_1(t) \leq d_1(0) + \sum_{j=1}^{t} Z_j \leq 2N + \sum_{j=1}^{t} Z_j
$$

or

$$
d_2(t) \leq d_2(0) - \sum_{j=1}^{t} Z_j \leq 2N - \sum_{j=1}^{t} Z_j
$$

holds. Set now $t = cn$ for some constant $c > 0$. Since $|d_i(t) - d_i(t-1)| \leq 1$ for any $i \in \{1, 2\}$ and any $t$, the robber is caught by the first cop by time $cn$, if $\sum_{j=1}^{cn} Z_j \leq -2N$, whereas in the second case he is caught by the second cop if $\sum_{j=1}^{cn} Z_j \geq 2N$. Since $\Pr(\sum_{j=1}^{cn} Z_j \geq 2N) = \Pr(\sum_{j=1}^{cn} Z_j \leq -2N)$, in both cases the robber is caught with probability at least the one that the simple random walk on the integers at time $cn$ is at a position bigger or equal to $2N$ (recall that the simple random walk is the standard random walk starting at the origin). Denote by $q_k(r)$ the probability that the simple random walk on the integers, starting at position $k \in \mathbb{N}$ has at least once reached the position $0$ by time $r$. By Theorem 2.17 of [30]

$$
q_k(r) \geq 1 - \frac{12k}{\sqrt{r}}.
$$

Applying the theorem with $k = 2N = 2\sqrt{n}(1 + o(1))$ and $r = cn$ we see that with probability at least $(1 + o(1))(1 - \frac{24}{\sqrt{c}})$ the simple random walk starting at $2N$ passes the origin by time $cn$. In other words, with at least this probability, there exists some time $r \leq 2N$ such that $\sum_{j=1}^{r} Z_j \geq 2N$. Conditioning under this event, by symmetry we get that with probability at least $\frac{1}{2}$, the value of $\sum_{j=1}^{cn} Z_j \geq 2N$. Thus, with probability at least

$$
c' = (1 + o(1))\frac{1}{2}\left(1 - \frac{24}{\sqrt{c}}\right),
$$

$d_i(cn) = 0$ for some $i \in \{1, 2\}$. This holds independently of the starting position of the robber, and thus, if the robber is not yet caught by time $cn$, the same lower bound holds for the probability of being caught in the next round of $cn$ steps. Hence, the expected number of rounds until the robber is caught is at most $\frac{1}{c'}$, and thus $dct_i(P_N \square P_N) \leq \frac{1}{c'}cn = O(n)$. $\qquad\square$

6.4. **Brooms.** In this section we consider a family of graphs which we call *brooms*. The broom graph $B(c, n)$, with $n$ vertices and a parameter $c$ $(0 < c \leq 1)$ is illustrated in Fig. 2. It consists of a path with $cn$ vertices, joined at one endpoint (the *center* of the broom) with a star of $(1 - c)n$ vertices. The *end* of the broom is the other endpoint of the path.

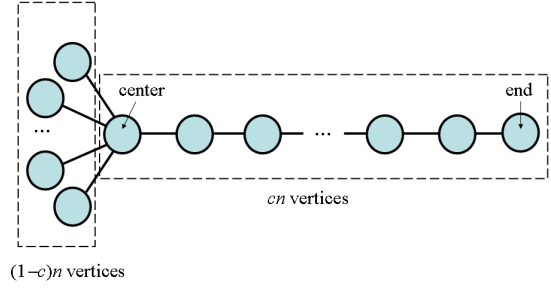

FIGURE 2. The broom graph $B(c, n)$.

Bounding $ct_i(B(c, n))$ and $dct_i(B(c, n))$ is interesting as an illustration of the game-theoretic approach. Perhaps more importantly, Corollary 6.10 shows that, for large $n$ and appropriately selected $c$, $F_i(B(c, n))$ can be arbitrarily close to any value in $[2, \infty)$. Note that Corollary 6.10 holds for every (sufficiently large) $n$, i.e., it is a property of the entire broom family.

**Theorem 6.8.** $dct_i(B(c, n)) = (1 + o(1))\frac{c^2 n}{2}$.

**Theorem 6.9.** $ct_i(B(c, n)) = (1 + o(1))n$.

**Corollary 6.10.** *For any $0 < c \leq 1$,*

$$F_i(B(c, n)) = \frac{ct_i(B(c, n))}{dct_i(B(c, n))} = (1 + o(1))\frac{2}{c^2}.$$

*Hence, for every $a \in [2, \infty)$ there exists $c \in (0, 1]$ such that $F_i(B(c, n)) = (1 + o(1))a$.*

*Proof of Theorem 6.8.* One possible strategy for the cop is the following: start at the center of the broom, wait there for one step, and then go to the end of the broom. Since the robber's position is distributed uniformly at random, and the robber has to move in every step, we have the following cases.

(1) With probability $1 - c$ the robber starts on a leaf of the star and is caught in $O(1)$ steps.
(2) With probability $c$ the robber starts on any vertex of the path and (starting at $t \geq 2$) the cop starts moving towards the end. In this case the problem is reduced to catching a drunk robber on a path of $cn$ vertices and this, by Theorem 6.1, takes $(1 + o(1))\frac{cn}{2}$ steps in expectation

Hence

$$dct_i(B(c, n)) \leq (1 - c) \cdot O(1) + c \cdot (1 + o(1))\frac{cn}{2} = (1 + o(1))\frac{c^2 n}{2}.$$

On the other hand, this is clearly the best strategy: with probability $c$ the robber is on one of the vertices of the path and, since the vertices of the star do not help, in this case, again by Theorem 6.1, the expected time to be caught is $(1 + o(1))cn/2$. Hence, $dct_i(B(c, n)) = (1 + o(1))\frac{c^2 n}{2}$. □

*Proof of Theorem 6.9.* For an upper bound we assign the following strategy for the cop. He starts at the end of the broom, moves to the center and, once there, he cheks all leaves in random order without repetitions. The robber's best response is to start at a randomly selected leaf and stay there until caught (actually he could move into a different leaf or even outside of the star as long as the cop is "sufficiently distant", but this will not change expected capture time). The expected capture time under these strategies is the sum of two terms:

(1) the cop needs $cn$ steps to go from the end to the center;
(2) once there, the cop will on the average need to check $(1-c)\,n/2$ leaves before he captures the robber and every such check requires two steps (one from the center to the leaf and one back), except the very last move which requires only one step.

Hence

$$ct_i\left(B\left(c,n\right)\right) \le cn + 2\left(1-c\right)n/2 = n.$$

A lower bound for $ct_i(B(c,n))$ can be established by describing a robber strategy and proving that it is optimal. Before we describe such a strategy we need some additional notation. Let us assign the "coordinate" 0 to the end of the broom and the "coordinate" $cn$ to the center. The cop's initial position $X_0$ can then be written as $bn$, where $b \in [0, c]$ (actually this excludes the possibility that $X_0$ is a leaf but, as we will soon see, we need not concern ourselves with this case). Note that $b$ is a parameter of the cop's strategy; also, the robber will observe $b$ before he makes his first move.

Before the game starts, the robber announces that he will use the following strategy: he will go to the end of the broom with probability $q = q(b) = b$, and to a randomly chosen leaf with probability $1-q$. The cop is aware of this and his only reasonable responses (after having started at $bn$) are the following.

(1) With probability $p$, the cop can go towards the end of the broom, and then back to the center to sweep all leaves.
(2) With probability $1-p$ he can go to the center, sweep a randomly chosen $x$-fraction of the leaves (for some $x \in [0,1]$), then move to the end of the broom, then go back to the center, sweeping all leaves (of course, for $x = 1$, he will have captured the robber by the time he reaches the end, at the latest; so he will not need to revisit the center).

We do not consider the possibility that the cop starts at a leaf because this does not influence the asymptotic behavior of the expected capture time; this justifies our previous claim that the cop's initial position can be parametrized by $bn$.

It is easily seen that the above *family* of strategies (parametrized by $(b, p, x)$) guarantees capture (of course capture may take place before the full schedule is executed). Note that if $c$ is exactly 1, we have a path, and $ct_i(B(c,n)) = (1 + o(1))n$. Also, if $b$ is exactly 0, the cop starts at the end of the broom, and the robber's only reasonable strategy is to hide at a randomly chosen leaf. In this case, the expected capture time is also $(1 + o(1))n$. Excluding these cases from the following analysis, we can break down the expected capture time into the following cases.

| R starts at | C moves towards | Prob. | Distance Travelled |
|---|---|---|---|
| end | end | $bp$ | $bn$ |
| a leaf | end | $(1-b)\,p$ | $(b + c + (1 - c))\,n$ |
| end | center | $b\,(1-p)$ | $((c-b) + 2x(1-c) + c)n$ |
| a leaf | center | $(1-b)\,(1-p)\,x$ | $((c-b) + x(1-c))\,n$ |
| a leaf | center | $(1-b)\,(1-p)\,(1-x)$ | $((c-b) + (2x(1-c) + 2c + (1-c))))\,n$ |

TABLE 1. The possible ways in which the robber can be captured.

The case in which the robber hides in the leaves and the cop starts by moving towards the center is broken down to two subcases (and hence requires two rows in the above table).

(1) In the first subcase, the cop checks an $x$-fraction of the leaves and captures the robber, *before* visiting the end of the broom. This happens with probability $x$ and the average number of leaves checked is $x \cdot (1 - c) \cdot n/2$.

(2) In the second subcase, the cop checks $x$-fraction of the leaves, fails to capture the robber, visits the end of the broom and then returns to check all the leaves (and so capture the robber with certainty). This happens with probability $1 - x$ and the number of leaves checked is $x \cdot (1 - c) \cdot n$ during the first visit and $(1 - c) \cdot n/2$ (on the average) during the second visit.

The expected capture time for a given value of $c$ is $E(T) = (1 + o(1))f_c(p, b, x) \cdot n$ and $f_c(p, b, x) \cdot n$ is obtained by multiplying the entries of the last two columns of Table 1, adding and performing some algebra. We finally get

$$
\begin{aligned}
f_c(b, p, x) &= (-1 + b + c + p - bc - bp - cp + bcp)\, x^2 + \\
&\quad (1 + b - 3c - p + bc - bp + 3cp - bcp)\, x + (2c - 2b + 2bp - 2cp + 1)\,.
\end{aligned}
$$

Hence $f_c(b, p, x)$ is a quadratic function in $x$, and (after some factorizations) can be rewritten as follows: $f_c(b, p, x) = a_2 x^2 + a_1 x + a_0$, with

$$
\begin{aligned}
a_2 &= -(1 - p)(1 - b)(1 - c) \\
a_1 &= (1 - p)(1 - 3c + bc + b) \\
a_0 &= 2(1 - p)(c - b) + 1.
\end{aligned}
$$

Since $a_2$ is negative (unless $p = 1$, where the whole function is zero), the parabola is downward concave, and thus it achieves its minimum either at $x = 0$ or at $x = 1$. If $x = 0$, $f_c(b, p, x) = a_0 = 1 + 2(1 - p)(c - b) \geq 1$. If $x = 1$, $f_c(b, p, x) = a_2 + a_1 + a_0 = 1$, and thus the minimum is achieved there. Thus, $ct_i(B(c, n)) \geq (1 - o(1))n$; this, together with the upper bound, shows that the robber strategy is optimal and that $ct_i(B(c, n)) = (1 + o(1))n$. $\qquad\square$

**Remark.** The broom $B(c, n)$ is a combination of the path $P_n$ and the star $S_N$: high $c$ values make $B(c, n)$ more like a path and low values more like a star. Accordingly, high $c$ values bring $F_i(B(c, n))$ closer to $F_i(P_n) = 2$ and low values increase $F_i(B(c, n))$ unboundedly, which is similar to the behavior of $F_i(S_N)$ for large $N$.

## 7. The "Infinite Speed" Robber

In this section we change the rules of the game slightly. Suppose the robber is still invisible but now is endowed with "infinite speed". In this case, what is iCOD?

We use the term "infinite speed" for brevity. What we actually assume is that the robber has speed $s \in \mathbb{N}$, i.e., during his turn he can traverse at most $s$ edges (as long as he does not go through a vertex containing a cop) and we examine what happens in the limit when $s$ tends to infinity. Let us note that the adversarial robber can choose to traverse fewer edges or even stay in place. As for the drunk robber, he simply performs $s$ steps of a random walk on $G$; if during one of these steps he runs into a cop, he is captured. Finally, the cops will still have unit speed.

We will use the notation $ct_i(G, s)$ to denote the expected capture time given that the robber is invisible, adversarial and has speed $s$; $dct_i(G, s)$ and $F_i(G, s)$ are defined similarly. Obviously we have $ct_i(G, 1) = ct_i(G)$, $dct_i(G, 1) = dct_i(G)$, $F_i(G, 1) = F_i(G)$. We are interested in the limits

$$
\lim_{s \to \infty} ct_i(G, s)\,, \qquad \lim_{s \to \infty} dct_i(G, s)\,, \qquad \lim_{s \to \infty} F_i(G, s)\,.
$$

In the case of drunk robber, it is actually quite easy to find $\lim_{s\to\infty} dct_i(G,s)$. Suppose C has $c(G)$ cops, and he keeps them stationary at some vertices $u_1, u_2, \ldots, u_{c(G)}$. As $s$ becomes large, the robber essentially performs a random walk on $G$ with $u_i$ ($i \in [c(G)]$) being absorbing vertices. Since a random walker visits every vertex of $G$ in finite expected time, the probability that the robber will hit an absorbing state during the first turn (conditional on *not* starting at absorbing state) approaches 1 as $s \to \infty$. Hence $\lim_{s\to\infty} dct_i(G,s) = \frac{n-c(G)}{n}$.

Next let us note that, for the adversarial robber, $s$ does not need to increase all the way to infinity. With $s = |V| = n$, the robber can in one turn reach any vertex in $G$. In other words $\lim_{s\to\infty} ct_i(G,s) = ct_i(G,n)$. In short,

$$\lim_{s\to\infty} F_i(G,s) = \frac{\lim_{s\to\infty} ct_i(G,s)}{\lim_{s\to\infty} dct_i(G,s)} = \frac{ct_i(G,n)}{\frac{n-c(G)}{n}} = \frac{n}{n-c(G)} ct_i(G,n).$$

Hence we now must study $ct_i(G,n)$. Let us first establish that $ct_i(G,n)$ exists. The game theoretic analysis of Section 4 holds for any value of $s$; a feasible robber strategy now specifies (probabilities on) moves which traverse at most $s$ edges without crossing a cop-occupied vertex. Hence $ct_i(G,n)$ is well defined.

It is easy to see that $\lim_{s\to\infty} F_i(G,s)$ can take any value in $\mathbb{N}$. For example, for the path $P_n$ we see immediately that $\lim_{s\to\infty} F_i(P_n,s) = \lim_{s\to\infty} \frac{n}{n-1} ct_i(P_n,s) = n$. As an additional example, let us consider $\lim_{s\to\infty} F_i(S_N,s)$ on the star $S_N$. When CiR is played on $S_N$, the adversarial infinite-speed robber can move to any leaf in his turn; hence he has more options than the unit-speed robber, who must remain in his original vertex for the entire game. However, let C control a single cop, who starts at vertex 0, searches a randomly selected leaf (with repetitions) at odd times and returns to 0 at even times. It is easy to see that this strategy is optimal and its expected capture time (i.e., $ct_i(G,s)$) is (for any $s \geq 2$):

$$ct_i(G,s) = \left( \sum_{t=1}^{\infty} \left( \frac{N-1}{N} \right)^{t-1} \frac{1}{N} (2t-1) \right) - 1 = 2N - 1.$$

(One time step is subtracted because after capture the cop does not need to return to vertex 0.) For the drunk robber we will have (for all $s$): $dct_i(G,s) = \frac{N}{N+1}$. Hence

$$\lim_{s\to\infty} F_i(S_N,s) = \frac{(2N-1) \cdot (N+1)}{N} = \Theta(n).$$

The example of the infinite speed robber on $S_N$ illustrates an additional interesting point. Recall the *graph search* (GS) game [42]. Similarly to CiR, GS involves a team of *searchers* (cops) and a *fugitive* (robber) who is invisible, adversarial and infinitely fast [19]. GS and CiR differ in one respect: in GS the fugitive is assumed to reside in the *edges*, not the vertices of $G$. But there are GS versions of *node search* and *mixed search*. Furthermore, in "classical" GS the cops are not restricted to move along edges; but there is a variant of GS, called *internal* search, in which this restriction is imposed. Let us simply state, without giving details, that the capture mechanism of CiR is equivalent to that of *internal mixed GS* (the interested reader can check this fact using [19] and the references therein).

We denote by $c_i^\infty(G)$ the minimum number of cops required to capture the invisible, adversarial and infinitely fast robber on $G$; we denote by $s(G)$ the minimum number of searchers required to guarantee capture of the fugitive in $G$, using internal mixed search. One would assume that these two games are equivalent and that $s(G) = c_i^\infty(G)$. However this is not the case: in the $S_N$ example we have $c_i^\infty(S_N) = 1 < 2 = s(S_N)$. Hence, at least for some graphs, we can have $c_i^\infty(G) < s(G)$. The reason for this is that in GS the fugitive is assumed to know in advance the *entire* itinerary of the searchers (i.e., he is *omniscient)*, while in CiR the robber only knows the past cop moves (and it suffices that the cops have a strategy that guarantees capture in finite time with probability one, not to have a strategy that catches the robber in finite number of steps.)

## 8. Conclusion

We have examined CiR, the invisible-robber version of the CR game. Our main interest was in the cost of drunkenness $F_i(G)$ which, as we have shown, exists for every graph $G$ and can get arbitrarily close to any value in $[2, \infty)$. To establish these results we have used concepts from game theory and the theory of partially observable Markov decision processes. For several families of graphs including stars, d-regular trees and grids we found (almost) exact bounds, and moreover we gave general upper and lower bounds. Finally, we have briefly examined the CiR variant in which the robber is *infinitely fast* and we have found, somewhat surprisingly, that the corresponding cop number $c_i(G)$ can be smaller than the (internal mixed) search number $s(G)$.

Our work can be extended in several ways. We first state some *open problems*.

(1) *For a general graph $G$ prove:* $|V(G)| \geq 2 \Rightarrow F_i(G) \geq 2$. We conjecture (but have not been able to prove) that this is true, because we have not found any graph $G$ with both $|V(G)| \geq 2$ and $F_i(G) < 2$.

(2) *For the square grid, find an asymptotically exact value of* $ct_i(P_N \square P_N)$. We have obtained a subexponential upper bound on $ct_i(P_N \square P_N)$ but we have not been able to find a nontrivial lower bound. By "nontrivial" we mean a tight bound; e.g., if a superpolynomial lower bound is found we will be very close to obtaining an asymptotically exact value for $ct_i(P_N \square P_N)$.

(3) As mentioned, $F_i(G)$ can get arbitrarily close to any value in $[2, \infty)$. On the other hand, there are values in $[2, \infty)$ which cannot be actually achieved by $F_i(G)$ (for example $F_i(G)$ cannot equal an irrational number). *Prove that: for any $a \in [2, \infty) \cap \mathbb{Q}$ there exists a $G$ such that $a = F_i(G)$.*

(4) *Find necessary and sufficient conditions for* $c(G) = c_i^\infty(G)$. The equality holds for paths, stars and cliques; for what other types of graphs does it hold?

(5) *Find necessary and sufficient conditions for* $c_i^\infty(G) = s(G)$, where $s(G)$ is the *internal mixed search number* of $G$.

Finally, we list several *future research directions*.

(1) We have obtained tight $F_i(G)$ bounds for paths, cycles, trees and grids. Can similar bounds be obtained for additional graph families (e.g., planar, bipartite, series-parallel, chordal, high girth, geometric)? What about random graphs?

(2) The use of *algorithms* becomes necessary for graphs which cannot be treated analytically. We want to develop *tractable* algorithms for the computation of $ct_i(G)$, $dct_i(G)$, $F_i(G)$ and also of optimal search strategies. It is well established in the literature that exact algorithms are computationally intractable; hence we intend to develop approximate algorithms, along with performance guarantees.

(3) We have studied the cost of drunkenness under a game theoretic model of the adversarial robber. As mentioned in Section 1, there is an alternative approach, in which the cops optimize their strategy under the assumption of an *omniscient* robber. We want to study the cost of drunkenness under this "*worst-case*" assumption.

(4) We have studied the cost of drunkenness, but our results can also be used to initiate a study (for both the adversarial and drunk variants) of the *cost of invisibility*. This can be expressed by the ratios $\frac{ct_i(G)}{ct(G)}$, $\frac{dct_i(G)}{dct(G)}$, the behavior of which we will study in the future.

**Acknowledgement:** We thank D. Thilikos (several discussions with him led us to the study of the CiR game) and A. Gurel-Gurevich (he provided help with Theorem 4.2).

DEPARTMENT OF MATHEMATICS, PHYSICS AND COMPUTER SCIENCES, ARISTOTLE UNIVERSITY OF THESSALONIKI, THESSALONIKI GR54124, GREECE
*E-mail address*: kehagiat@auth.gr

DEPARTMENT OF MATHEMATICS, RYERSON UNIVERSITY, TORONTO, ON, CANADA, M5B 2K3
*E-mail address*: dmitsche@ryerson.ca

DEPARTMENT OF MATHEMATICS, RYERSON UNIVERSITY, TORONTO, ON, CANADA, M5B 2K3
*E-mail address*: pralat@ryerson.ca